\documentstyle[aps,epsfig,twocolumn]{revtex}
\begin{document}
\draft
\title{Universal phase boundary shifts for corner wetting and filling} 

\author{A.\ O.\ Parry$^1$, A.\ J.\ Wood$^1$, E.\ Carlon$^2$ and A.\ Drzewi\'nski$^3$}
\address{
$^1$Department of Mathematics, Imperial College, London SW7 2BZ, United
Kingdom\\ 
$^2$INFM, Dipartimento di Fisica, Universit\`a di Padova, I-35131
Padova, Italy
$^3$ Institute of Low Temperature and Structure Research, Polish Academy
of Sciences, P.O.Box 1410, 50-950 Wroc\l aw 2, Poland\\ 
}
\maketitle

\begin{abstract}
The phase boundaries for corner wetting (filling) in square and diagonal 
lattice Ising models are exactly determined and show a universal shift 
relative to wetting near the bulk criticality. More generally, scaling theory 
predicts that the filling phase boundary shift for wedges and cones is 
determined by a universal scaling function $R_d(\psi)$ depending only on 
the opening angle $2\psi$. $R_d(\psi)$ is determined exactly in $d=2$
and approximately in higher dimensions using non-classical local
functional and mean-field theory. Detailed numerical transfer matrix
studies of the the magnetisation profile in finite-size Ising squares
support the conjectured connection between filling and the
strong-fluctuation regime of wetting.\\  
\end{abstract}
            
The wetting of structured and chemically heterogeneous substrates has 
recently drawn a great deal of interest
\cite{Dietrev,LandL,Gau,Michelle,Carlos2,MFT,OurPRL1,OurPRL2,OurJphysC,JphysCone}. This is motivated not only by the possible relevance to
emerging technologies such as micro-fluidics \cite{micro1}, but also from a
more fundamental statistical physics perspective since surface structure
may induce new types of interfacial phase transition. Examples of these
are morphological phenomena on heterogeneous substrates
\cite{LandL,Gau} and filling transitions for fluids adsorbed near
corners, wedges and cones
\cite{MFT,OurPRL1,OurPRL2,OurJphysC,JphysCone}. Recent continuum effective
interfacial Hamiltonian studies indicate that the conditions for
observing continuous wedge and cone filling are much more relaxed than
for continuous (critical) wetting. These studies also indicate that
fluctuation effects at filling are characterised by a high degree of
universality which persists even in the presence of long-ranged forces. For example, the filling of a three dimensional wedge is 
characterised by a universal roughness exponent \cite{OurPRL2}
whilst for two dimensional corners (in both ordered and disordered
systems \cite{OurPRL1,OurJphysC}) and three dimensional cones
\cite{JphysCone} fluctuation effects mimic, quite
precisely, behaviour predicted to occur for the strong-fluctuation (SFL)
regime of critical wetting \cite{lipFish}. 

Here we go beyond the effective Hamiltonian approach
and investigate surface phase diagrams and 
fluctuation effects at filling using microscopic theory. We 
present exact results \cite{next} for two dimensional ($d=2$) Ising models which
improve upon previous low temperature, solid-on-solid (SOS) studies
\cite{DandO,ALip} and 
show that, near the bulk critical point, the
shift of the phase boundary (relative to wetting) is universal, determined only
by the corner opening angle. The results of detailed numerical transfer-matrix 
studies, which test the proposed connection between two dimensional filling and 
the SFL, are discussed at length and predictions for phase boundary shifts for 
wedges and cones in $d=3$ are made.

Consider a two dimensional corner with opening angle $2 \psi$ (with $\psi\in
(0,\pi/2]$) in contact with a
bulk phase $A$ which is at two-phase bulk $AB$ coexistence. Thus  
$\psi=\pi / 4$ is a right angle corner whilst $\psi=\pi / 2$ is a flat, 
semi-infinite 
interface. Thermodynamic arguments \cite{Finn} indicate that the corner is
completely filled by phase $B$ when the contact angle $\theta$ satisfies 

\begin{equation} 
\theta(T) = \frac{\pi}{2} - \psi
\label{one}
\end{equation}
implying that filling precedes wetting. The same
condition applies for the filling of three-dimensional wedges and cones and is
confirmed by effective Hamiltonian studies \cite{MFT,OurPRL1,OurPRL2}. Now
recall that, for systems with sufficiently 
short-ranged forces, the scaling of the excess surface-free energy near the bulk
critical temperature $T_c$ \cite
{NakFish} implies that $\theta = W(ah_1t^{-\Delta_1})$, where $t =
(T_c - T)/T_c$ is the reduced temperature, $h_1$ is the surface field and
$\Delta_1$ the surface gap exponent. Note that $W(x)$ can be identified as a
universal scaling function provided we allow for a non-universal metric
factor $a$. It follows that near the bulk critical point
the filling phase boundary satisfies  $ah_1^f(T)t^{-\Delta_1} = 
W^{-1}(\frac{\pi}{2} - \psi)$ generalising the well
known result for the wetting phase boundary (denoted $h_1^w(T)$)
corresponding to 
$\psi=\pi/2$ \cite{NakFish}. Consequently the phase boundary shift defined as  
\begin{equation}
\lim_{T\to T_c}\frac{h_1^f(T)}{h_1^w(T)}=R_d (\psi)  
\label{two}
\end{equation}
should be universal depending only on the opening angle and dimensionality. 
Implicit in this 
definition is the assumption that both the wetting and filling
transitions are continuous. This requirement is not restrictive in the two
dimensional systems, which are the main concern of this paper, since both the 
filling and wetting transitions are continuous at all temperatures. In three
dimensions  we emphasise that both wedge and cone filling may be continuous even if the 
wetting transition is first-order. To continue, it follows that for
fixed, small values of the surface field
\begin{equation}
\frac{T_c-T_w}{T_c-T_f}=R_d(\psi)^{1/\Delta_1}
\label{three}
\end{equation}
which provides a
useful means of estimating $T_f$ provided $T_w$ is known. Moreover, whilst 
the character of the filling transitions for the 
$d=3$ wedge and cone are different, the location of the phase boundary 
obeys the same thermodynamic condition (\ref{one}) and the shift 
function $ R_3 (\psi)$ is unique. 

To test the scaling theory we consider square and diagonal lattice right-angle 
Ising corners which are natural extensions of Abraham's semi-infinite model
\cite{Doug1980,AKoS}(see Fig.1). We assume isotropic interactions in order to preserve reflection
symmetry about the apex diagonal so that both boundaries behave as
identical walls. The reduced interaction strength between nearest
neighbour spins is $K (=1/T)$, except they are modified to $\tilde{K}
(=h_1/T)$ in the bottom row and far left column. The boundary spins are
fixed to +1 whilst the bulk magnetisation (infinitely far from the
boundaries) is supposed negative.

\begin{figure}[h]
\label{figCorner}
\vspace*{0.5cm}
\centerline{\epsfig{file=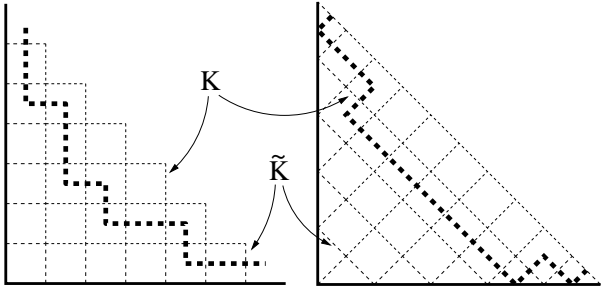,width=8.cm}}
\vspace*{0.5cm}
\caption{Square (S) and diagonal (D) lattice isotropic 
Ising corners with
weakened bonds. Typical low-temperature SOS interfacial configurations are shown.}
\end{figure}
  
 These boundary conditions induce a
long domain wall which rounds the corner at some
distance $L$ (measured along the bottom row or left column) from the apex. On 
approaching the filling transition the mean value of $L$ and hence the 
diagonal interfacial height $z$, measured in the \texttt{[11]}
direction from the corner apex, diverge.  
Defining the incremental free-energy $\tau_p$ for each
semi-infinite lattice as the difference of the wall-down spin and
wall-up spin surface tensions recall that for square (S) 
and diagonal (D) Ising lattices \cite{Doug1980,AKoS}
\begin{equation}
\cosh{\tau_p} = \cosh{\tau} - \frac{1}{2} F(T)
\label{DougBasic}
\end{equation}
where $\tau^{S}=2K +\ln \tanh K $ and 
$\tau^{D}=2 \ln (\sinh 2K)$ denote 
the free interfacial tensions. The singular contributions to the free energy are

\vspace*{0.2cm}
\begin{eqnarray}
F^D (T) & = & \frac{(e^K (\cosh 2 K - \cosh 2 \tilde{K}) - e^{-K} \sinh
2 K)^2}{(\cosh 2 K - \cosh 2 \tilde{K}) \sinh 2K} \nonumber \\
F^S (T) & = & \frac{(\sinh^2 2K - 1 - 2 \sinh^2 2
\tilde{K})^2}{\sinh^2 2K}
\label{four}
\end{eqnarray}
which vanish at wetting. Next note that for large $L$ the excess
free-energy of the corner must contain a term $f(T)L$ arising from
surface tension contributions with $f^S(T)=\tau_p^S -\tau^D/2$ and
$f^D(T)=\tau_p^{D}-\tau^{S}$. At filling $f(T_f)=0$, leading to the exact
conditions  
\begin{equation}
\cosh 2 \tilde{K} = \cosh2K -e^{-2K}\sinh^22K 
\label{5}
\end{equation}
\begin{equation}
\sinh^2 2 \tilde{K} = \frac{(\sinh 2K - 1)(1 - e^{-2K})}{2}
\label{6}
\end{equation}
for the square and diagonal lattices respectively (see Fig. 2). Both types of
corner show the same universal phase boundary shift with
\begin{equation}
R_2(\pi/4)= \frac{1}{\sqrt{2+\sqrt{2}}}
\label{7}
\end{equation}
in agreement with scaling theory. The low temperature behaviour is also
revealing. For the square lattice there 
\begin{figure} \label{Plots}
\vspace*{0.5cm}
\centerline{\epsfig{file=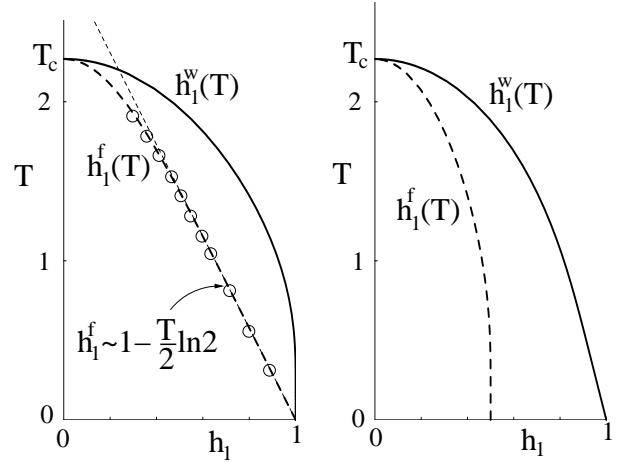,width=8.cm}}
\vspace*{0.5cm}
\caption{Surface phase diagrams for the filling in the square and
diagonal (shown on the left and right respectively) lattice Ising
corners compared with critical wetting. The circles on the left hand 
figure show the numerically determined loci where the magnetisation
profile shows finite-size scaling behaviour characteristic of the SFL 
regime of wetting}
\end{figure}
is a strong entropic effect arising from the energetic degeneracy of domain
configurations and away from the bulk critical regime we find
\begin{equation}
h_1^{f} \sim 1 - \frac{T}{2} \ln 2
\label{SOS}
\end{equation}
which is precisely the SOS result obtained earlier by Duxbury and
Orrick \cite{DandO} and Lipowski \cite{ALip}. For the diagonal lattice
on the other hand such entropic effects influence the wetting but not filling
phase boundary. The SOS limit for this lattice has not been considered
before but is very easily constructed. Assuming the interface sticks to the 
walls (as shown) outside of the filled region the probability distribution for
diagonal interfacial height follows as
\begin{equation}
P(z)= \frac{e^{-z/\langle z\rangle}}{\langle z\rangle}
\label{9}
\end{equation}
where $\langle z\rangle \sim (h_1-h_1^f)^{-1}$ and $\lim_{T\to 0}h_1^f=1/2$
true to the exact Ising result. Notice that the scaling of $P(z)$ 
is precisely the same as that predicted for the SFL regime of
critical wetting in agreement with continuum 
effective Hamiltonian studies of filling in open wedges \cite{OurPRL1}. 

Returning to the Ising phase boundary, we note that the above results allow us 
to determine the exact $R_2(\psi)$ even
though it is not always possible to (easily) construct Ising lattices with arbitrary
opening angle. Near the bulk critical temperature the Ising model recovers
fluid isotropy and the incremental free-energy $\tau_p$ may be 
identified with $\tau \cos\theta$. Using the thermodynamic 
condition for filling (\ref{one}) we obtain, after a little manipulation, the elegant expression
\begin{equation}
R_2 (\psi) = \sqrt{2} \sin \frac{\psi}{2}
\label{R2}
\end{equation}
which recovers the result for the right-angle corner quoted above as a special
case. Note that for a
corner with $120^\circ$ opening angle we predict 
$R_2(\pi/3)=1/\sqrt2$ , which may be tested using the triangular lattice Ising model which 
naturally forms corners of this type. Also (\ref{R2}) is consistent with
the general requirement for the open wedge ($\psi\to\pi/2$) limit
$1-R_d(\psi)\sim(\pi/2-\psi)^{2/2-\alpha_s}$  which follows
from (\ref{one}). Here $\alpha_s$ 
denotes the critical wetting specific heat exponent which is zero for
the planar Ising model \cite{Doug1980}.

\begin{figure}
\label{first}
\vspace*{0.5cm}
\centerline{\epsfig{file=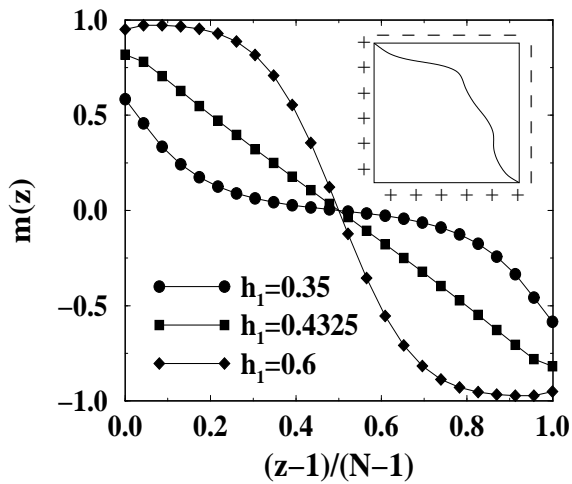,width=8cm}}
\vspace*{0.5cm}
\caption{Numerical transfer matrix results for the magnetisation
profile in the finite-size Ising square for $N=24$, $T=1.6$ and different surface
fields above, below and (nearly) at the filling phase boundary.}
\end{figure}

To test the connection between filling and the SFL regime for the full Ising
model we consider a finite-size square lattice Ising square of side length $N$
with weakened bonds around the perimeter. The boundary spins are fixed to +1
(-1) along the bottom (top) rows and left (right) column respectively so that
the domain wall runs from the upper left to lower right corner (see inset 
Fig.3). In the limit of $N$ tending to infinity the system decouples into two separate
corner lattices, each of which exhibit a filling transition by the opposite bulk
phase at the same temperature, $T_f$. For finite $N$ this anti-symmetric
choice of boundary conditions frustrates the interface analogous to
behaviour in infinite Ising strips with opposing surface fields
\cite{Albano,PE}. Far above the
filling transition the interface stretches from the
top left to bottom right corners across the middle of the square and has an
 r.m.s
width of order $\sqrt N$. In contrast, far below the transition temperature we anticipate 
pseudo phase coexistence, since the interface may be bound to either the
top-right or the lower-left corner. The cross-over between these different types of interfacial
behaviour occurs when $T_f-T\sim N^{-1/\beta_0}$ reflecting the influence of length-scales associated with the filling transition.

 The conjectured connection between filling and SFL wetting suggests that at 
$T=T_f$
the magnetisation profile along the diagonal $m(z)$ is universal and characterised
by the same scaling function describing the profile for an
infinite Ising strips with opposing surfaces fields at $T=T_w$. Thus exactly
at $T=T_f$ and for asymptotically large $z,N$ with arbitrary $z/N\in(0,1)$, we 
expect the SFL-like scaling behaviour \cite{PEN,MS}
\begin{equation}
m(z) = m_0 \left( 1 - \frac{2z}{N} \right)
\label{seven}
\end{equation}
where $m_0$ is the bulk spontaneous magnetisation. Notice that at the filling
temperature the r.m.s. width of the interface now scales with $N$ compared to
$\sqrt N$ for $T\geq T_f$ which reflects the additional excitations
present near filling \cite{JphysCone}. Using numerical transfer
matrix techniques \cite {next} we have calculated the exact magnetisation profile in the 
Ising square for systems up to $N=24$. The results verify the qualitative 
character of the interface delocalisation discussed above and are in excellent
quantitative agreement with the scaling of the profile at $T_f$. Indeed, testing for the most linear profile
turns out to be a highly efficient numerical method of determining the phase 
boundary up to $T=1.6$ (see Fig.2) beyond which the influence of the bulk 
critical point becomes apparent and
larger systems sizes are required to see the scaling of $m(z)$. At low
temperatures the exact Ising results are very well described by the SOS
approximation 
and for this model we have numerically computed $m(z)$ for systems up to $N=180$. Again 
there is
excellent agreement with the scaling prediction (11) for the profile at $T=T_f$
and the SOS phase boundary $h_1^f(T)$ is recovered to four significant figures.

 It is straightforward to derive an expression for $R_d$ 
 using mean-field (MF) theory. This serves not only as point of comparison with the
exact Ising result but also indicates the $d$ dependence. We omit the details 
\cite {next} and only quote the final result (which is independent of 
$d$) obtained using Landau theory \cite{NakFish}
\begin{equation}
\sin\psi=\frac{3}{2}R(1-\frac{R^2}{3})
\label{forR}
\end{equation}
which we anticipate to be correct for $d>4$. This is a 
remarkably good approximation to the exact $d=2$ result (yielding 
$R(\pi/4)=0.518$ and $R(\pi/3)=0.684$) and strongly suggests that the dimension 
dependence of $R_d(\psi)$ is rather weak. Unfortunately no $\epsilon$ expansion results 
for the pertinent surface tensions are available to systematically estimate 
$R_d$ in lower dimensions; though progress can be made using approximate non-classical local-functional theory 
which is usually a
reliable treatment of bulk and surface criticality. Strictly speaking
the theory fails for $d\leq3$ when $\psi$ is very close to $\pi/2$ since it 
does not account for capillary-wave like fluctuations which alter the Landau 
MF value of $\alpha_s$ (equal to zero). However, the asymptotic critical regime
for wetting in three dimensions, where such effects are important, is extremely
small and the value of $R_3(\psi)$ will be uninfluenced by capillary-waves for 
all but the shallowest of wedges. The local functional theory we
use is a simple extension and application of the Fisk-Widom theory \cite {FW}. Again we
omit the details of the calculation and only quote the result for three
dimensions \cite{next} which is conveniently written as the expansion
\begin{equation}
\sin \psi \approx \frac{7R_3}{2\sqrt6}(1-\frac{R_3^2}{4}-\frac{9R_3^4}{160}+..)
\label{FW}
\end{equation}
valid provided $R_3$ is not extremely close to unity. For the  
right-angle wedge the local theory predicts $R_3(\pi/4)\approx 0.53$ which lies close to and
between the $d=2$ and $d=4$ (MF) results and is presumably reliable to within 
a few per cent. This may be used to estimate the location of the filling phase
boundary for Ising model right 
angle wedges with weak surface fields which are natural generalisations of
models used to study wetting \cite{BinderRev}. Again, as in $d=2$, there are two ways of 
constructing this wedge 
with a simple cubic lattice with cross sections the same as shown in Fig.1
(and translational invariance in the other spatial direction). Away from the 
critical region we anticipate that, for the analogue of the (S) lattice,
entropic effects are important and
 the phase boundary is determined by a linear law similar to (\ref{SOS}). This 
prior knowledge of the phase boundary may be
useful when studying the large scale fluctuation effects predicted for wedge
filling characterised by
the universal critical exponent $\beta_0=1/4$ \cite{OurPRL2}. 
 
 In summary we have presented exact results for corner wetting in $d=2$ and
approximate results in higher dimensions, highlighting the universal shift of
the phase boundary relative to wetting in the bulk critical region. 
In $d=2$ numerical transfer matrix
studies of the finite-size scaling of the magnetisation profile in Ising 
squares strongly support the conjectured relation between filling and the SFL 
regime. 

AJW acknowledges support from the EPSRC and AD thanks the Polish Academy of
Sciences and the Royal Society for Travel Grants.
 

\begin{references}
\bibitem{next} Further details of our analytical and numerical results will 
appear elsewhere.
\end{references}
\end{document}